\documentclass[letter,twocolumn,a4paper]{jpsj3}

\usepackage{amsmath,amssymb,amsfonts}
\usepackage{graphicx}
\usepackage{bm}

\title{Renormalization Effects on Quasi-Two-Dimensional Organic Conductor
$\alpha$-(BEDT-TTF)$_2$I$_3$}

\author{\name{Hiroki \surname{Isobe}}$^1$ and 
	\name{Naoto \surname{Nagaosa}}$^{1,2}$}
\inst{$^1$
	Department of Applied Physics, University of Tokyo, Tokyo 113-8656, Japan \\
$^2$Cross Correlated Materials Research Group (CMRG) and Correlated
Electron Research Group (CERG), ASI, RIKEN, Wako, Saitama 351-0198, Japan}

\abst{The quasi-two-dimensional organic semiconductor 
$\alpha$-(BEDT-TTF)$_2$I$_3$ 
[BEDT-TTF=bis(ethylenedithio)tetrathiafulvalene]
has an anisotropic linear dispersion with a zero energy gap near the Fermi level. 
Owing to the vanishing density of states at the Fermi level, the Coulomb interaction 
is unscreened in this material. We theoretically study the effect of the long-range 
Coulomb interaction and the low-energy/long-wavelength behavior of
$\alpha$-(BEDT-TTF)$_2$I$_3$ using the renormalization group analysis. 
The nearly logarithmic enhancement of the 
velocity reshapes tilted Dirac cones, and changes the low-temperature behavior. 
We also show the theoretical calculation for the site-selective spin susceptibility, 
which can be measured in an NMR experiment. }

\kword{organic conductor, $\alpha$-(BEDT-TTF)$_2$I$_3$, tilted Dirac cone, 
renormalization group}

\begin{document}
\maketitle

\begin{figure}
\centering
\includegraphics[width=0.9\hsize]{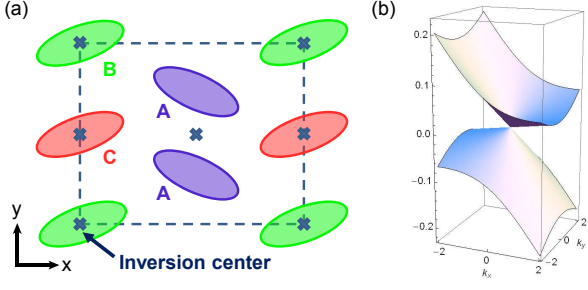}
\caption{(Color online) 
(a) Lattice structure of the conducting BEDT-TTF layer. A, B, and C denote the 
inequivalent sites. The inversion centers are also depicted.  (b) Effective energy 
dispersion near the tilted Dirac cone ($\xi=+1$). The Dirac cone is largely tilted in the 
$x$-direction. The units in the graph are \AA$^{-1}$ for momentum and eV for energy.}
\label{fig:lattice}
\end{figure}

Salts of the organic molecule BEDT-TTF [bis(ethylenedithio)tetrathiafulvalene] 
show various electronic phases, such as 
Mott insulators, charge transfer insulators, semimetals, and superconductors~\cite{seo}.  
In addition, there are many types of crystal structures.  
$\alpha$-(BEDT-TTF)$_2$I$_3$ is one such organic compound 
\cite{bender,tajima2009,kobayashi2009}, and consists of alternately stacked 
BEDT-TTF molecule layers and tri-iodide (I$_3^-$) anion layers.  
BEDT-TTF molecular planes constitute conducting layers, on which a 
quasi-two-dimensional electronic system is formed.  
The symmetry of the crystal structure of $\alpha$-(BEDT-TTF)$_2$I$_3$ 
is low with only the spatial inversion.  The unit cell contains four BEDT-TTF molecules, 
three of which are crystallographically inequivalent
[Fig. \ref{fig:lattice}(a)].  

The band calculation predicts that 
$\alpha$-(BEDT-TTF)$_2$I$_3$ has a semimetallic Fermi surface at 
room temperature and ambient pressure~\cite{seo}.  
The first-order phase transition to a charge-ordered insulating phase~\cite{kino2} 
occurs at about 135\,K with marked changes in susceptibility~\cite{rothaemel} and 
resistivity~\cite{tajima2006}. 
At higher pressures, this charge-ordered insulating phase will gradually be suppressed 
and will completely vanish at 1.5\,GPa~\cite{tajima2006}.  
It has been revealed that there exists a state 
with an anisotropic linear dispersion near the Fermi level, by the band calculation 
\cite{kino,katayama2006,kobayashi2007}.  
According to this calculation, $\alpha$-(BEDT-TTF)$_2$I$_3$ has two strongly tilted 
Dirac cones, and the tilting is caused by 
the nearest-neighbor and next-nearest-neighbor hoppings.  
Experimental results, e.g., the $T^2$ dependence of the carrier density 
\cite{tajima2006,kajita1992}, are consistent with this linear dispersion.  

Graphene~\cite{novoselov,zhang} is a representative material with Dirac cones.  
Unlike $\alpha$-(BEDT-TTF)$_2$I$_3$, graphene has a purely two-dimensional electron 
system with isotropic Dirac cones.  The band crossing point is just at the Fermi level, 
and the vanishing density of states (DOS) leads to an unscreened long-range 
Coulomb interaction.  
This has been analyzed by the renormalization group (RG) approach, 
and the Dirac cone reshaping due to the renormalized Fermi velocity 
is observed by Shubnikov-de Haas oscillations as the change 
in the cyclotron frequency~\cite{elias}.  

Compared with that of graphene, 
the narrow bandwidth of $\alpha$-(BEDT-TTF)$_2$I$_3$ 
strengthens the electron correlation effect.  
It leads to the enhancement of Dirac cone reshaping, and its effects are 
expected to be observed more easily than the effect of graphene. 
In this study, we analyze the long-range Coulomb interaction effect on the system 
with tilted Dirac cones using the RG approach, and calculate the spin susceptibility 
as a physically measurable quantity. 

We start with the following generalized Weyl Hamiltonian describing the tilted massless
Dirac cones~\cite{goerbig}:
\begin{equation}
\mathcal{H}(\bm{k}) = \xi\bm{w}\cdot\bm{k} + v_x k_x \sigma_x + v_y k_y \sigma_y,
\end{equation}
where $\xi=\pm 1$ denotes the valley degeneracy and we set $\hbar=1$. 
For the moment, we consider the $\xi = +1$ valley. 
The parameter $\bm{w}$ determines the tilt of the anisotropic Dirac cone.
The energy of this model [Fig. \ref{fig:lattice}(b)] is 
\begin{equation}
E_\pm(\bm{k}) = \bm{w}\cdot\bm{k} \pm \sqrt{v_x^2 k_x^2 + v_y^2 k_y^2}.
\end{equation} 
We assume that the parameters satisfy the relation
\begin{equation}
\left(\dfrac{w_x}{v_x}\right)^2 + \left(\dfrac{w_y}{v_y}\right)^2 < 1.
\end{equation}
This condition ensures that the system has a point node. 

We consider the anisotropic long-range Coulomb interaction 
\begin{equation}
V(\bm{q}) = \frac{2\pi e^2}{\varepsilon \sqrt{q_x^2 + \eta q_y^2}}
\end{equation}
as a perturbation to the system. 
The anisotropy of the dielectric constant is reflected in the factor $\eta$, and 
$\varepsilon$ is the dielectric constant.
The unperturbed Green's function is 
\begin{equation}
G_0(\bm{k}, \omega) = 
\frac{1}{\omega - \bm{w}\cdot\bm{k} - v_x k_x \sigma_x - v_y k_y \sigma_y}.
\end{equation}

The RG analysis of two-dimensional systems often treats Coulomb interaction 
with large-$N$ expansion~\cite{gonzalez,son}. 
In the method for the isotropic system, the Coulomb propagator 
$D_0(\bm{k},\omega)$ is modified by adding a one-loop fermion bubble diagram 
with $N$ fermion species: 
\begin{equation}
D_0(\bm{k},\omega) = \left( 2|\bm{k}| + \frac{Ne^2}{8\varepsilon}
	\frac{\bm{k}^2}{\sqrt{v^2\bm{k}^2-\omega^2}} \right)^{-1}.
\end{equation}
The dressed term has importance in the strong coupling limit, but in the weak 
coupling case, it gives only a small correction to the result. 
We concentrate our analysis on the low-temperature region, where the running 
coupling constant becomes smaller than the bare value, 
so the dressed term is neglected in the following analysis.   

\begin{figure}
\centering
\includegraphics[width=0.4\hsize]{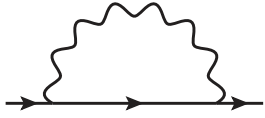}
\caption{One-loop self-energy considered for RG analysis.}
\label{fig:diagram}
\end{figure}

With the RG analysis, the parameters $\bm{v}$ and $\bm{w}$ are 
modified by the electron self-energy $\Sigma(\bm{k},\omega)$.
We calculate the self-energy to one-loop order.
The one-loop-order self-energy $\Sigma^{(1)}(\bm{k},\omega)$ 
[Fig. \ref{fig:diagram}] is evaluated by
\begin{equation}
\Sigma^{(1)}(\bm{k},\omega) = i \int \frac{d\omega'}{2\pi} \frac{d^2 p}{(2\pi)^2}
	G_0(\bm{p},\omega+\omega') V(\bm{k}-\bm{p}).
\end{equation}
The momentum integral is taken in the momentum shell 
$\Lambda e^{-l} \leq |\bm{p}| \leq \Lambda$. 
After some calculation, we obtain 
\begin{equation}
\Sigma^{(1)}(\bm{k},\omega) = 
	\frac{\alpha_x}{4}v_x l k_x \sigma_x 
	 + \frac{\alpha_y}{4}v_y l k_y \sigma_y, 
\end{equation}
where the coupling constants $\alpha_x$ and $\alpha_y$ are defined by
\begin{subequations}
\begin{align}
\alpha_x\! = \frac{4e^2}{\varepsilon \pi}\! \int_0^{\frac{\pi}{2}} \!\!
	\frac{\cos^2\theta d\theta}{(v_x^2\cos^2\theta + v_y^2\sin^2\theta)^{\frac{1}{2}}
		(\cos^2\theta + \eta\sin^2\theta)^{\frac{3}{2}}}, \\
\alpha_y\! = \frac{4e^2}{\varepsilon \pi}\! \int_0^{\frac{\pi}{2}} \!\!
	\frac{\sin^2\theta d\theta}{(v_x^2\cos^2\theta + v_y^2\sin^2\theta)^{\frac{1}{2}}
		(\cos^2\theta + \eta\sin^2\theta)^{\frac{3}{2}}}.
\end{align}
\end{subequations}

Then, the RG equations for $v_x$ and $v_y$ are
\begin{subequations}
\label{eq:rg}
\begin{align}
\frac{dv_x}{dl}=&\frac{\alpha_x}{4}v_x, \\
\frac{dv_y}{dl}=&\frac{\alpha_y}{4}v_y.
\end{align}
\end{subequations}
By setting $v_x=v_y$ and $\eta=1$, these formulae reduce to the isotropic case, 
like graphene~\cite{kotov}.
At the one-loop level of self-energy, the tilting parameter $\bm{w}$ is not renormalized 
and stays constant. 
The discussions above are unchanged for the $\xi = -1$ valley.  

\begin{figure}
\centering
\includegraphics[width=0.9\hsize]{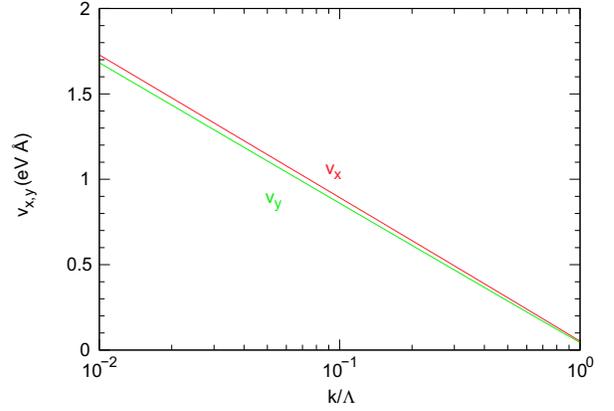}
\caption{(Color online)
Numerical solutions to the RG equations. Both $v_x$ and $v_y$ show 
almost completely logarithmic dependences on the momentum scale.} 
\label{fig:velocity}
\end{figure}

The numerical solutions to eq.~\eqref{eq:rg} are shown in Fig.~\ref{fig:velocity}.
The initial values at the cutoff momentum are $v_x=0.0515, v_y=0.0439, 
w_x=-0.0389$, and $w_y=0.0048$ (in eV\,\AA)~\cite{katayama2009}, and we set 
$\varepsilon=10$ and $\eta=1$. 
$v_x$ and $v_y$ show nearly logarithmic dependences on the momentum scale, 
as in the system with the isotropic Dirac cone.

\begin{figure}
\centering
\includegraphics[width=0.8\hsize]{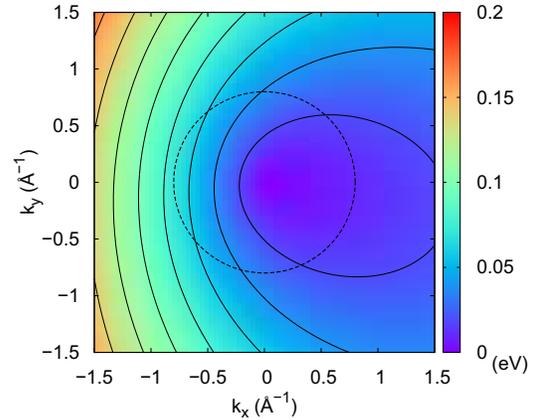}
\caption{(Color online) 
Energy distribution in the $k$-plane in the noninteracting case. 
The cutoff circle $\Lambda = 0.8\,\mathrm{\AA}^{-1}$ is shown by the dashed line. 
The solid lines denote the constant-energy curves, plotted every 0.02\,eV. 
The gradient of the cone is low near $\theta\sim0$ and high near $\theta\sim\pi$.
The energy on the cutoff circle is not constant owing to the tilting of the Dirac cone.}
\label{fig:energy}
\end{figure}

The site-selective spin susceptibility $\chi_\alpha$ ($\alpha$ = A, B, and C) 
is described by~\cite{katayama2009}
\begin{equation}
\chi_\alpha = \int_{-\infty}^\infty d\epsilon D_\alpha(\epsilon) 
	\left( -\frac{\partial f}{\partial \epsilon} \right),
\end{equation}
where $D_\alpha(\epsilon)$ is the site-dependent DOS, 
and $f(\epsilon)$ is the Fermi distribution. 
The definition of the site-dependent DOS is
\begin{equation}
D_\alpha(\epsilon) = 2\int\frac{d^2k}{(2\pi)^2} |d_\alpha(\theta)|^2 
	\delta(\epsilon-E_+(\bm{k})),
\end{equation}
where $d_\alpha(\theta)$ represents the eigenstate for each site, 
and the absolute values are given by
\begin{subequations}
\begin{align}
|d_\text{A}(\theta)|^2 &= 0.270\cos^2(\theta/2) + 0.195\sin^2(\theta/2), \\
|d_\text{B}(\theta)|^2 &= 0.610\sin^2(\theta/2), \\
|d_\text{C}(\theta)|^2 &= 0.460\cos^2(\theta/2).
\end{align}
\end{subequations}
The site-dependent DOS reflects the $\theta$ dependence of the eigenstates.  
The DOS of site C mainly comes from the gentle slope of the tilted Dirac cone 
($\theta\sim 0$), 
and the DOS of site B comes from the steep slope ($\theta\sim\pi$) 
[Fig.~\ref{fig:energy}]. 
Site A has almost no angular dependence. 

Site C, which corresponds to the gentle slope of the tilted Dirac cone, has the 
largest contribution to the spin susceptibility, because it has the highest density 
of states among the three sites.
In contrast, site B has the lowest spin susceptibility. 
Although the site-dependent DOS leads to a different magnitude of the site-selective
spin susceptibility, the DOS for each site is proportional to the energy $\epsilon$. 
We can easily see that $\chi_\alpha \propto T$ in the noninteracting system.

For a two-dimensional system with a linear dispersion, the DOS 
$D(\epsilon)$ is proportional to the energy $\epsilon$.  However, the enhancement of 
the velocities revealed by the RG analysis changes the energy dispersion, 
and sufficiently below 
the cutoff energy, the density of states as a function of the energy is suppressed 
compared with the noninteracting one.  This scheme is valid for 
$\alpha$-(BEDT-TTF)$_2$I$_3$ within the temperature range where the dispersion is 
well approximated by a linear one. 

\begin{figure}
\centering
\includegraphics[width=0.9\hsize]{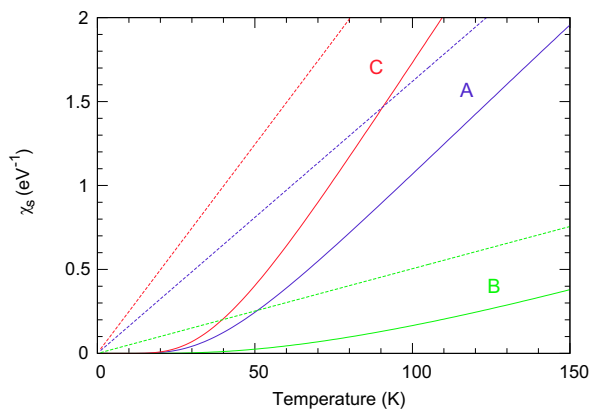}
\caption{(Color online)
Theoretical calculation of the site-selective spin susceptibility. 
The solid and dashed lines describe the spin susceptibility for the interacting and 
noninteracting cases, respectively.} 
\label{fig:nmr}
\end{figure}

The theoretical result for the site-selective spin susceptibility is obtained 
by numerical calculation and is shown in Fig.~\ref{fig:nmr}. 
We set the cutoff $\Lambda = 0.8\,\mathrm{\AA}^{-1}$. 
Compared with the noninteracting result, which shows a linear dependence of 
$\chi_\alpha$ in $T$, 
the RG analysis reveals a reduction in the spin susceptibility 
$\chi_\alpha$ at low temperatures.
The validity of the linear dispersion approximation also depends on the angle $\theta$. 
The temperature ranges where the linear dispersion approximation holds are 
$T\lesssim 70$\,K for the gentle slope and $T\lesssim 100$\,K for the steep slope.

Another important behavior is that the characteristic temperature of 
the site-selective spin susceptibility $\chi_\alpha$ is different for each site.
Here, the characteristic temperature indicates the point where the spin susceptibility 
rapidly grows.  
Although we set the circular cutoff momentum, the energy 
at the cutoff depends on the momentum direction owing to the tilting of 
the Dirac cone [Fig.~\ref{fig:energy}]. 
This fact leads to a strong suppression of the site-B spin susceptibility. 

In summary, we have studied the effect of the long-range Coulomb interaction in 
a system with tilted Dirac cones using perturbative RG analysis. 
The velocity enhances logarithmically, as observed in an isotropic case 
such as graphene.
We calculate the site-selective spin susceptibility for the quasi-two-dimensional 
organic conductor $\alpha$-(BEDT-TTF)$_2$I$_3$. 
The RG analysis indicates a reduced the site-selective spin susceptibility 
at low temperatures, 
and the characteristic temperatures are different for each site.

We thank M. Hirata and K. Kanoda for fruitful discussion. 
This work is supported 
by a Grant-in-Aid for Scientific Research (Grant No. 24244054)
from the Ministry of Education, Culture,
Sports, Science and Technology of Japan, 
by Strategic International Cooperative Program (Joint Research Type)
from the Japan Science and Technology Agency, 
and by Funding Program for World-Leading Innovative RD on Science and
Technology (FIRST Program).

\end{document}